\documentclass[a4paper]{jpconf}
\usepackage{graphicx}
\usepackage{subfigure}  
\begin{document}
\title{Recent Results of Fluctuation and Correlation Studies from the STAR Experiment}

\author{Terence J. Tarnowsky}

\address{National Superconducting Cyclotron Laboratory, Michigan State University, East Lansing, MI 48824, USA}

\ead{tarnowsk@nscl.msu.edu}

\begin{abstract}
Enhanced fluctuations and correlations have been observed in the phase transitions of many systems. Their appearance at the predicted QCD phase transition (especially near the expected critical point) may provide insight into the nature of the phase transition. Recent results from the STAR experiment will be presented with a focus on particle ratio (K/$\pi$ and p/$\pi$) fluctuations, forward-backward multiplicity correlations, and balance functions. Also discussed will be possibilities for measuring these correlations and fluctuations in the recently initiated beam energy scan at RHIC.
\end{abstract}

\section{Introduction}

Fluctuations and correlations are well known signatures of phase transitions. Specifically, the quark/gluon to hadronic phase transition may lead to significant fluctuations \cite{Koch1}. As part of an ongoing beam energy scan at the Relativistic Heavy Ion Collider (RHIC), the search for the QCD critical point will make use of the study of correlations and fluctuations, particularly those that could be enhanced during a phase transition that passes close to the critical point. The STAR experiment has studied a number of correlation and fluctuation observables as a function of energy, system-size, and centrality. Results from the STAR experiment on particle ratio fluctuations ($K/\pi$ and $p/\pi$), forward-backward multiplicity correlations, balance functions, and net charge fluctuations are discussed.

\section{Measurements}
\subsection{Particle Ratio Fluctuations}

Dynamic particle ratio fluctuations, specifically fluctuations in the $K/\pi$ and $p/\pi$ ratio, can provide information on the quark-gluon to hadron phase transition \cite{Strangeness1, Strangeness2, Strangeness3}. These can be measured using the variable $\nu_{dyn}$, originally introduced to study net charge fluctuations \cite{nudyn1, nudyn2}. $\nu_{dyn}$ quantifies deviations in the particle ratios from those expected for an ideal statistical Poissonian distribution. The definition of $\nu_{dyn,K/\pi}$ (describing fluctuations in the $K/\pi$ ratio) is,

\begin{eqnarray}
\nu_{dyn,K/\pi} = \frac{<N_{K}(N_{K}-1)>}{<N_{K}>^{2}}
+ \frac{<N_{\pi}(N_{\pi}-1)>}{<N_{\pi}>^{2}}
- 2\frac{<N_{K}N_{\pi}>}{<N_{K}><N_{\pi}>}
\label{nudyn}
\end{eqnarray}

where $N_{K}$ and $N_{\pi}$ are the number of kaons and pions in a particular event, respectively. In this proceeding, $N_{K}$ and $N_{\pi}$ are the total charged multiplicity for each particle species. A formula similar to (\ref{nudyn}) can be constructed for $p/\pi$ fluctuations. $\nu_{dyn}$ = 0 for the case of a Poisson distribution of kaons and pions and is largely independent of detector acceptance and efficiency in the region of phase space being considered \cite{nudyn2}. An in-depth study of $K/\pi$ fluctuations in Au+Au collisions at $\sqrt{s_{NN}}$ = 200 and 62.4 GeV was carried out by the STAR experiment \cite{starkpiprl}.

Previous measurements of particle ratio fluctuations utilized the variable $\sigma_{dyn}$ \cite{NA49}, 

\begin{equation}
\sigma_{dyn} = sgn(\sigma_{data}^{2}-\sigma_{mixed}^{2})\sqrt{|\sigma_{data}^{2}-\sigma_{mixed}^{2}|}
\label{signudyn}	
\end{equation}

where $\sigma$ is the reduced width of the $K/\pi$ or $p/\pi$ distribution in either real data or mixed events. It has been shown that $\nu_{dyn}$ and $\sigma_{dyn}$ are related as,

\begin{equation}
\sigma_{dyn}^{2} \approx \nu_{dyn}
\label{signudyn2}
\end{equation}

\begin{figure}
\centering
\subfigure[]{
\includegraphics[scale=0.40]{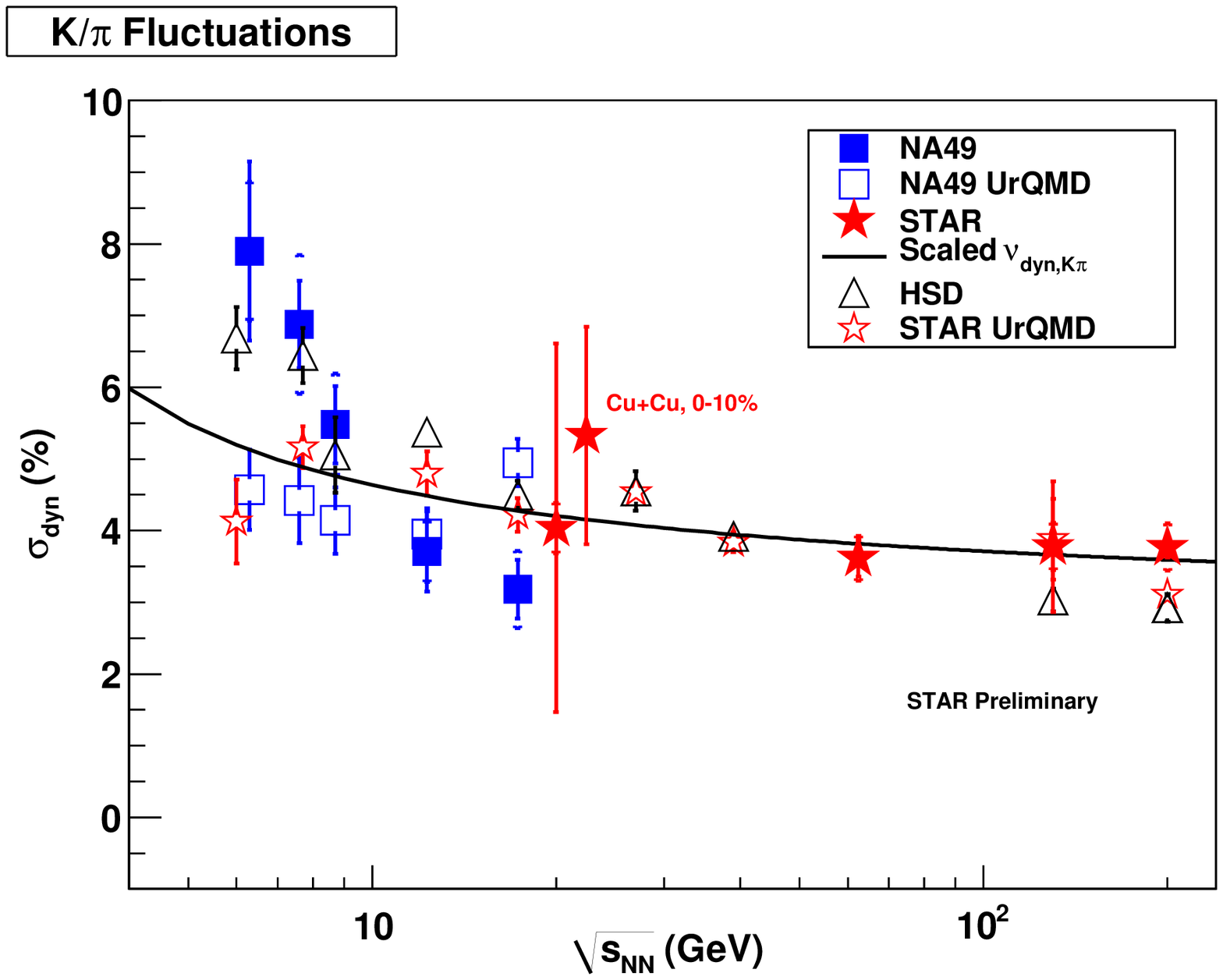} 
\label{kpi_excitation}
}
\subfigure[]{
\includegraphics[scale=0.40]{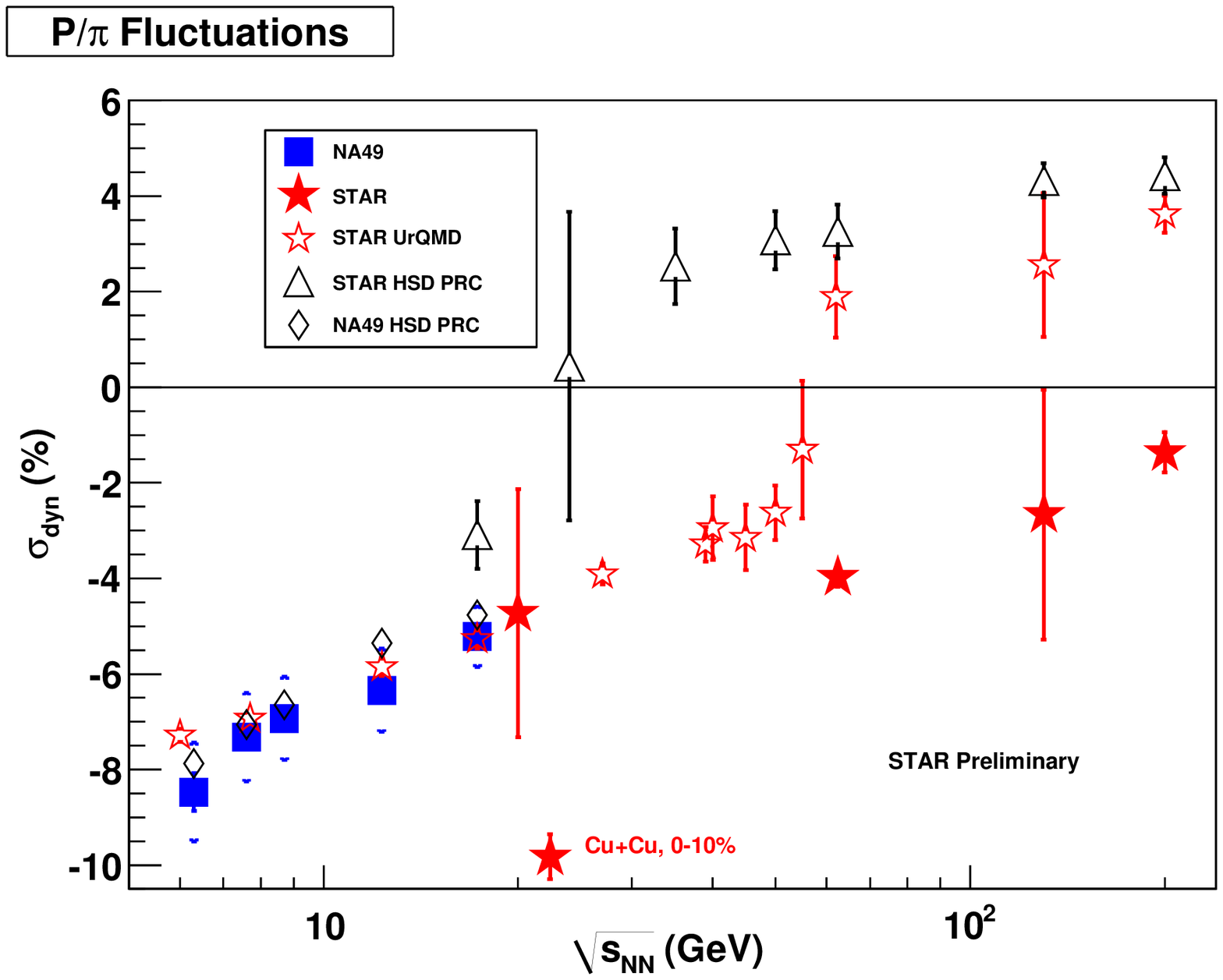} 
\label{ppi_excitation}
}
\caption{$K/\pi$ (left) and $p/\pi$ (right) fluctuations expressed as $\sigma_{dyn}$, as a function of incident energy. Data is shown from both the NA49 (solid blue squares) and STAR experiments (solid red stars) from central collisions, Pb+Pb, 0-3.5\% for NA49 and Au+Au, 0-5\% for STAR, except for the Cu+Cu point at $\sqrt{s_{NN}}$ = 22.4 GeV (0-10\%). Also shown are model calculations from UrQMD using the NA49 (open blue squares) and STAR (open red stars) experimental acceptances and HSD (open black triangles) with the STAR and NA49 (open black diamonds) acceptance. Models labeled ``STAR'' are Au+Au collisions, while models labeled ``NA49'' are Pb+Pb. The solid black line in Figure \ref{kpi_excitation} shows the relationship between $\nu_{dyn}$ at high energies converted to $\sigma_{dyn}$ and extrapolated to lower energies. Errors include both statistical and systematic effects, except for the STAR point in Cu+Cu at $\sqrt{s_{NN}}$ = 22.4 GeV, which has only statistical errors.}
\label{asdf}
\end{figure}

The excitation function for $K/\pi$ fluctuations expressed as $\sigma_{dyn}$ is shown in Figure \ref{kpi_excitation}. To convert $\nu_{dyn}$ to $\sigma_{dyn}$, the relationship in (\ref{signudyn2}) was used. Errors were also propagated from $\nu_{dyn}$ to $\sigma_{dyn}$. There is a strong decrease with increasing incident energy for the NA49, Pb+Pb 0-3.5\% results (solid blue squares). The STAR results (solid red stars) for 0-5\% Au+Au at $\sqrt{s_{NN}}$ = 19.6, 62.4, 130, and 200 GeV, and 0-10\% Cu+Cu at $\sqrt{s_{NN}}$ = 22.4 GeV all show the same value of $\sigma_{dyn}$ within errors. $\sigma_{dyn}$ is approximately flat above $\sqrt{s_{NN}}$ = 10 GeV for all systems. Also shown are model predictions for UrQMD processed for the NA49 (open blue squares) and STAR (open red stars) experimental acceptances, respectively. A third model prediction, Hadron String Dynamics (HSD) \cite {HSD} processed using the STAR experimental acceptance (open black triangles) is also plotted.

Figure \ref{ppi_excitation} shows the excitation function for $p/\pi$ fluctuations expressed using the variable $\sigma_{dyn}$. The general trend demonstrated by the data is smaller negative fluctuations with increasing incident energy for both NA49, Pb+Pb 0-3.5\% results (solid blue squares) and STAR results (solid red stars) for 0-5\% Au+Au collisions. The outlier data point is from 0-10\% Cu+Cu collisions at $\sqrt{s_{NN}}$ = 22.4 GeV, which is much more strongly negative. Both UrQMD with the STAR experimental acceptance (open red stars), and HSD with the NA49 and STAR experimental acceptances (open diamonds and triangles, respectively, reproduce most of the low energy data and over predict the high energy data points.

\subsection{Forward-Backward Multiplicity Correlations}

Forward-backward multiplicity correlations have been studied in elementary particle collisions \cite{LRCee,LRCee2,LRCee3,LRCpp,LRCpp2,LRCpp3,LRCpp4}, and more recently, from the STAR experiment in heavy ion collisions at RHIC \cite{fb_prl}. FB correlations are measured as a function of pseudorapidity ($\eta$). Because they probe the longitudinal characteristics of the system produced in heavy ion collisions, they provide unique insight into the space-time dynamics and the earliest stages of particle production. A relationship between the multiplicity in the forward and backward hemisphere was found in hadron-hadron collisions \cite{fblinear},

\begin{equation}\label{linear} 
<N_{b}(N_{f})> = a + bN_{f} 
\end{equation}

where the coefficient {\it b} is referred to as the correlation strength. Eq. \ref{linear} can be expressed in terms of the expectation values,

\begin{equation}\label{dispersion}
b = \frac{<N_{f}N_{b}>-<N_{f}><N_{b}>}{<N_{f}^{2}>-<N_{f}>^{2}} = \frac{D_{bf}^{2}}{D_{ff}^{2}}
\end{equation}

where $D_{bf}^{2}$ and $D_{ff}^{2}$ are the backward-forward and forward-forward dispersions, respectively. All quantities in Eq.~(\ref{dispersion}) are measurable experimentally. The presence of long-range FB correlations has been predicted in some particle production models, specifically the Dual Parton Model (DPM) and the Color Glass Condensate/Glasma (CGC) model \cite{DPM, CGC}. Both models contain a form of longitudinal color flux tubes: color strings in the DPM and Glasma flux tubes in the CGC \cite{DPM_CGC}. These longitudinal structures are predicted to be the origin of the long-range correlation.

\begin{figure}
\centering
\includegraphics[width=100mm]{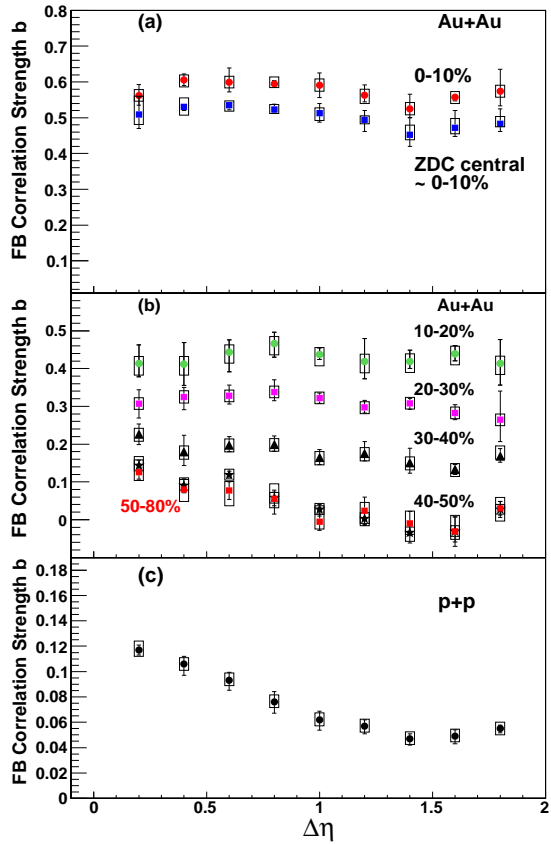}
\caption{The centrality dependence of the FB correlation strength for Au+Au at $\sqrt{s_{NN}}$ = 200 GeV for central 0-10\%, 10-20\%, 20-30\%, 30-40\%, 40-50\%, and peripheral 50-80\%. The long-range correlations are seen in the most central Au+Au data, but behave similar to that of only short-range correlations beginning at 30-40\% most central data. The correlation strength in the most peripheral Au+Au collisions resembles that of \textit{pp}.}
\label{FB_Fig}
\end{figure}

Figure \ref{FB_Fig} shows the forward-backward (FB) correlation strength ({\it b}) as a function of centrality from $\sqrt{s_{NN}}$ = 200 GeV Au+Au collisions. From most central (0-10\%) to peripheral (50-80\%) collisions there is a general decrease in the FB correlation strength. The shape of the FB correlation strength for the 50-80\% bin as a function of $\Delta\eta$ indicates the presence of mostly short-range correlations, which are expected to decrease exponentially with increasing $\eta$ gap. If only short-range correlations contributed to the correlation, all centralities would resemble the 50-80\% result. The presence of a strong correlation that grows with centrality and is flat beyond $\Delta\eta > 1.0$ is indicative of a long-range correlation produced by multiparton interactions in central $\sqrt{s_{NN}}$ = 200 GeV Au+Au collisions.

\subsection{Balance Function}

\begin{figure}
\centering
\includegraphics[width=80mm]{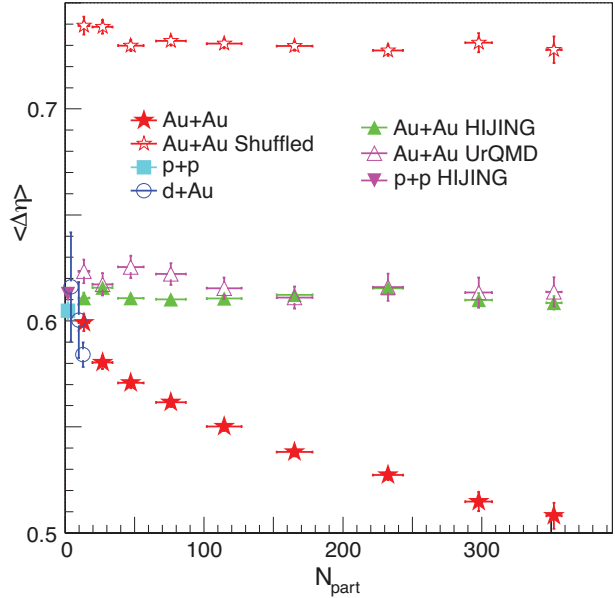}
\caption{The centrality dependence of the width of the inclusive charged particle balance function ($<\Delta\eta>$) in $\sqrt{s_{NN}}$ = 200 GeV Au+Au collisions, compared to model predictions and to \textit{pp} and d+Au collisions at the same energy. The narrowing of the balance function width in central Au+Au collisions may be an indication of delayed hadronization.}
\label{BF_fig}
\end{figure}

The charge balance function has been predicted to be sensitive to the creation of a deconfined system of quarks and gluons by indicating a delay in the time to hadronization \cite{BF}. If hadrons are produced locally in pairs of oppositely charged particles, those produced early in the collision will experience more interactions (such as scattering and/or diffusion). Those pairs produced later will experience less of these effects. The width of the balance function is controlled by this, as well as the relative momentum between the produced pair, which is dictated by the thermal properties of the system. This leads to a prediction of a wide balance function for early production of hadrons and a narrow balance function for delayed hadronization. Delayed hadronization can result from a deconfined system of quarks and gluons that prevents the production of hadrons for some amount of time. The balance function can be expressesd as,

\begin{equation}
B(\Delta\eta)=\frac{1}{2}\left\{\frac{N_{+-}(\Delta\eta)-N_{++}(\Delta\eta)}{N_{+}}+\frac{N_{-+}(\Delta\eta)-N_{--}(\Delta\eta)}{N_{-}}\right\}
\label{BFeq}
\end{equation}

where $N_{xy}$ (x,y = + or -) is the number of charged particle pairs in the pseudorapidity range, $\Delta\eta = \eta_{2}-\eta{1}$. The terms in Equation \ref{BFeq} are calculated from pairs on an event-by-event basis and the results are summed over all events. The width is quantified using the weighted average, $<\Delta\eta>$. Figure \ref{BF_fig} shows the width of the balance function in $\Delta\eta$ for inclusive charged particles, for \textit{pp}, d+Au, and Au+Au collisions at $\sqrt{s_{NN}}$ = 200 GeV. For Au+Au and d+Au, it is shown as a function of centrality ($N_{part}$) and for \textit{pp} only minimum bias data is shown. The Au+Au data is compared to two models: HIJING and URQMD. Also shown are the results from shuffled Au+Au events. Shuffled events remove the charge correlation among particles in that event. The result from event shuffling demonstrates the widest width of the balance function that is possible inside the finite detector acceptance. Figure \ref{BF_fig} shows that the width of the balance function peripheral Au+Au collisions at $\sqrt{s_{NN}}$ = 200 GeV is in good agreement with the width in \textit{pp} and d+Au collisions at the same energy. The width of the balance function decreases as a function of increasing centrality in Au+Au collisions, and is narrowest in the most central (top 0-5\%) collisions. This effect is not reproduced by the models, which demonstrate a constant balance function width as a function of centrality. . This is consistent with the idea of delayed hadronization in central $\sqrt{s_{NN}}$ = 200 GeV Au+Au collisions due to the formation of a partonic system.

\section{Outlook}

Beginning in 2010 RHIC embarked on a Beam Energy Scan (BES) program, motivated primarily by the possibility of finding evidence of a QCD critical point, as shown by lattice QCD predictions \cite{LQCD_CP,LQCD_CP2,LQCD_CP3}. The current program will be split over two years and cover a range in energies from $\sqrt{s_{NN}}$ = 200 GeV down to as low as $\sqrt{s_{NN}}$ = 5 GeV. The STAR detector is currently in a configuration with a minimum amount of material between the beam pipe and the main detector, the Time Projection Chamber (TPC). With large acceptance in $\eta$ and $\phi$ and a fully installed Time-of-Flight detector, STAR is fully prepared to study fluctuations and correlations at all energies, while searching for signatures of the QCD critical point.

\section{Summary}

Several different fluctuation and correlation measurements from the STAR experiment have been discussed. Large changes in event-by-event fluctuation and correlation observables are expected in the hadron to quark-gluon phase transition, and also near the critical point. STAR continues to search for rapid changes of these observables as a function of incident collision energy and centrality. In the next two years, STAR will have large statistics datasets, acquired data across a range in energies from $\sqrt{s_{NN}}$ = 200 to $\sqrt{s_{NN}}$ = 5 GeV. This spans a broad area of the QCD phase diagram, including that of a predicted QCD critical point, though experimental limitations limit study to the one particular value of $\mu_{B}$ for each chosen collision energy. The study of fluctuations and correlations (including those not discussed here), are expected to be primary signatures of the proposed QCD critical point accessible at RHIC.
\section*{References}

\bibliography{all}
\bibliographystyle{unsrt}


\end{document}